\documentclass[10pt,conference]{IEEEtran}

\usepackage{graphicx}      %
\usepackage{booktabs}      %
\usepackage{amsmath}
\usepackage{amssymb}

\usepackage{color}

\usepackage{xspace}
\newcommand{\tool}{\textsc{EvoVuln}\xspace}

\usepackage{cite}   %

\usepackage{enumitem}

\usepackage{listings}
\usepackage{xcolor}

\definecolor{codegreen}{rgb}{0,0.6,0}
\definecolor{codegray}{rgb}{0.5,0.5,0.5}
\definecolor{codepurple}{rgb}{0.58,0,0.82}
\definecolor{backcolour}{rgb}{0.95,0.95,0.92}

\lstdefinestyle{mystyle}{
    backgroundcolor=\color{backcolour},   
    commentstyle=\color{codegreen},
    keywordstyle=\color{magenta},
    numberstyle=\tiny\color{codegray},
    stringstyle=\color{codepurple},
    basicstyle=\ttfamily\footnotesize,
    breakatwhitespace=false,         
    breaklines=true,                 
    captionpos=b,                    
    keepspaces=true,                 
    numbers=left,                    
    numbersep=5pt,                  
    showspaces=false,                
    showstringspaces=false,
    showtabs=false,                  
    tabsize=2
}

\usepackage{multirow}
\usepackage{multicol}

\usepackage{colortbl}
\newcommand{\cv}[1]{%
  \ifnum#1>89\cellcolor{gray!72}\else%
  \ifnum#1>79\cellcolor{gray!58}\else%
  \ifnum#1>69\cellcolor{gray!45}\else%
  \ifnum#1>59\cellcolor{gray!33}\else%
  \ifnum#1>49\cellcolor{gray!22}\else%
  \ifnum#1>39\cellcolor{gray!13}\else%
  \ifnum#1>0\cellcolor{gray!6}\else%
  \cellcolor{white}%
  \fi\fi\fi\fi\fi\fi\fi #1%
}

\newcommand{\etal}{\textit{et al.}\xspace}
\usepackage{hyperref}      %
\usepackage{cleveref}
\crefname{section}{\S}{\S\S}
\Crefname{section}{Section}{Sections}

\usepackage[skip=2pt]{caption}
\newcommand{\distance}{2pt}
\begin{document}

\title{Knowledge Over Parameters: Evolving Smart Contract Vulnerability Detection}

\author{
\IEEEauthorblockN{Yuqiang Sun\IEEEauthorrefmark{1},
Han Liu\IEEEauthorrefmark{2},
Ying Li\IEEEauthorrefmark{3},
Yiran Zhang\IEEEauthorrefmark{1},
Zong Cao\IEEEauthorrefmark{4},
Ziyun Guo\IEEEauthorrefmark{5},
Yang Liu\IEEEauthorrefmark{1}}
\IEEEauthorblockA{\IEEEauthorrefmark{1}Nanyang Technological University, Singapore\\
yuqiang.sun@ntu.edu.sg, yiran002@e.ntu.edu.sg, yangliu@ntu.edu.sg}
\IEEEauthorblockA{\IEEEauthorrefmark{2}Nankai University, China\quad hanliu@nankai.edu.cn}
\IEEEauthorblockA{\IEEEauthorrefmark{3}University of California, Los Angeles, USA\quad ying.li@ucla.edu}
\IEEEauthorblockA{\IEEEauthorrefmark{4}Imperial Global Singapore, Imperial College London, and Nanyang Technological University\quad z.cao@imperial.ac.uk}
\IEEEauthorblockA{\IEEEauthorrefmark{5}Singapore Management University, Singapore\quad zyguo@smu.edu.sg}
}

\maketitle

\begin{abstract}
Smart contract vulnerabilities are predominantly logic bugs whose detection requires structured, step-by-step procedural knowledge of attack patterns and contract semantics. Existing LLM-based methods struggle to generate this knowledge automatically: prompt-based methods rely on manually crafted detection rules, while fine-tuning requires massive labeled datasets that are inherently scarce in this domain. We present \tool, an automated framework that reformulates vulnerability detection as a procedural knowledge evolution problem, synthesizing and refining detection logic using only a minimal number of labeled samples.
To achieve this, \tool introduces two key mechanisms. First, a Runtime with an Inversion of Control (IoC) architecture compiles detection rules into Executable Policies. This strictly decouples deterministic control flow from LLM semantic reasoning, ensuring faithful logical adherence and producing dense diagnostic telemetry for precise error localization. Second, a two-phase evolution pipeline refines the rule via abductive semantic debugging without any parameter updates: \textit{Cold Start} bootstraps and stress-tests an initial rule using auto-synthesized corner cases; \textit{Few-Shot Evolving} then grounds the policy in real-world semantics using only five vulnerable and five safe examples per vulnerability type.

Evaluated across five real-world vulnerability types, \tool achieves a 71\% macro-average F1-score, outperforming all baselines. The evolved procedural knowledge is portable across models: it enables a lightweight, low-cost model to surpass a much larger zero-shot model by 19 percentage points, and transfers to other LLMs without retraining, at a one-time evolution cost under \$50.
\end{abstract}

\section{Introduction}
\label{sec:introduction}

Smart contracts are high-stakes programs governing decentralized finance (DeFi), where logic bugs, such as price manipulation and access control flaws, frequently lead to massive financial losses~\cite{defihacks2026}. Unlike syntactic errors, these logic flaws depend on semantic intent, requiring structured, multi-step procedural knowledge to verify if a contract violates application-level invariants.

Current detection paradigms face a fundamental bottleneck. 
Prompt-based methods rely on manually authored rules~\cite{sun2024gptscan,liu2025propertygpt,ding2025smartguard}, which are labor-intensive and fail to scale to emerging threats. 
Conversely, fine-tuning approaches~\cite{ma2025combining,jie2025agent4vul,wang2024smartinv} attempt to learn detection patterns directly, yet are constrained by the severe scarcity of labeled data. 
Neither paradigm supports the automated generation or evolution of detection logic.

Inspired by the agentic ``skill'' paradigm, which encodes reusable, task-specific capabilities to avoid reasoning from scratch~\cite{wang2025re,jiang2026sok,schick2023toolformer}, we view vulnerability detection as the synthesis of procedural knowledge: structured rules that dictate semantic inspection steps. Existing methods like GPTScan~\cite{sun2024gptscan} essentially instantiate this paradigm with human-authored rules, but they leave three critical gaps:

First, current procedural knowledge relies entirely on human experts, making it unscalable for emerging vulnerability types. 
When relying on LLMs to generate this knowledge automatically, the generated procedures are often flawed or insufficient to cover real-world attack surfaces. 
Second, even if correct knowledge is provided, when unconstrained LLM agents attempt to autonomously execute these complex procedures, they suffer from reasoning drift, frequently falling back on their own internal priors instead of faithfully executing the prescribed multi-step logic.
Conversely, bypassing procedural logic through direct fine-tuning is intractable due to the severe scarcity of labeled data. This raises the central question of this paper: \textit{Can procedural detection knowledge be automatically generated, strictly enforced, and continuously evolved using minimal labeled data, without manual engineering?}

We present \tool, a framework that answers this question affirmatively by reformulating vulnerability detection as an automated lifecycle of procedural knowledge generation and evolution. \tool treats procedural knowledge not as a static prompt, but as a persistent software artifact that is synthesized, executed, and empirically refined. To operationalize this framework, we address three primary dimensions:  
(\textit{i}) \textit{Controlled Execution via Inversion of Control (IoC)}: Even perfectly generated procedural knowledge is useless if the agent fails to follow it faithfully. \tool compiles detection rules into Executable Policies, deterministic programmatic detection plans that strictly enforce structural control flow, utilizing LLMs solely as on-demand semantic oracles for localized judgments.
(\textit{ii}) \textit{Abductive Knowledge Evolution}: Automatically generated procedural knowledge is prone to errors and requires feedback grounded in actual contract behavior. \tool employs abductive semantic debugging to iteratively patch and evolve detection rules. By leveraging dense execution telemetry from the executable policy, \tool pinpoints logical missteps in failed detections and performs targeted updates using minimal labeled samples.
(\textit{iii}) \textit{Knowledge Transfer Across Models}: \tool's evolved procedural knowledge is portable: it enables lightweight, low-cost models (e.g., GPT-5-nano) to outperform much larger models in zero-shot mode, decoupling detection accuracy from model scale.

We evaluate \tool on five vulnerability types against static analysis~\cite{feist2019slither}, rule-based~\cite{sun2024gptscan}, deep learning-based~\cite{nguyen2022mando}, fine-tuning-based~\cite{yu2025sael,ma2025combining}, zero-shot LLMs~\cite{gpt522026,gpt5nano2026}, and coding agent baselines~\cite{claudecode,codex}.
\tool achieves a macro-average F1-score of 71\%, substantially outperforming all baselines.
Notably, even when utilizing the lightweight GPT-5-nano as the detector, a model significantly weaker than GPT-5.2, \tool outperforms GPT-5.2 zero-shot by 19 percentage points, demonstrating that structured procedural knowledge combined with controlled execution can compensate for model capability gaps.
Furthermore, the evolved rules are highly portable: transferring them to Qwen3.5-9B and MiniMax-M2.5 without any retraining yields macro F1-score of 66\% and 67\% respectively, both substantially above GPT-5.2 zero-shot (52\%), confirming that the knowledge itself rather than the underlying model drives the performance gains.
The total API cost for evolving rules across all five vulnerability types is under \$50, and the evolution process is one-time: once a rule converges, it can be deployed directly without repeating the training process.

In summary, this paper makes the following contributions:
\begin{itemize}
  \item We reformulate smart contract logic vulnerability detection as a \textit{procedural knowledge generation and evolution} problem, and propose \tool, an automated framework that generates and refines detection procedures through a semantic lifecycle.
  \item We introduce an IoC execution architecture, \textit{Runtime}. By compiling procedural knowledge into Executable Policies, we decouple deterministic control flow from LLM semantic reasoning, ensuring faithful execution and providing dense diagnostic telemetry.
  \item We propose a highly data-efficient, dual-phase evolution mechanism: \textit{Cold Start} validates initial knowledge via synthesized corner cases, while Few-Shot Evolving performs abductive semantic debugging using minimal real-world examples.
  \item We demonstrate that \tool outperforms nine baselines, showing that evolved procedural knowledge enables lightweight models to surpass top-tier LLMs used in zero-shot mode.
\end{itemize}

\section{Background \& Motivation}
\label{sec:motivation}

As highlighted in~\cref{sec:introduction}, LLM-based contract vulnerability detection encounters three bottlenecks: the severe data scarcity that hinders fine-tuning, the unreliability of LLM-generated rules, and the reasoning drift of agents during execution.

\begin{figure}[t]
    \centering
    \includegraphics[width=\linewidth]{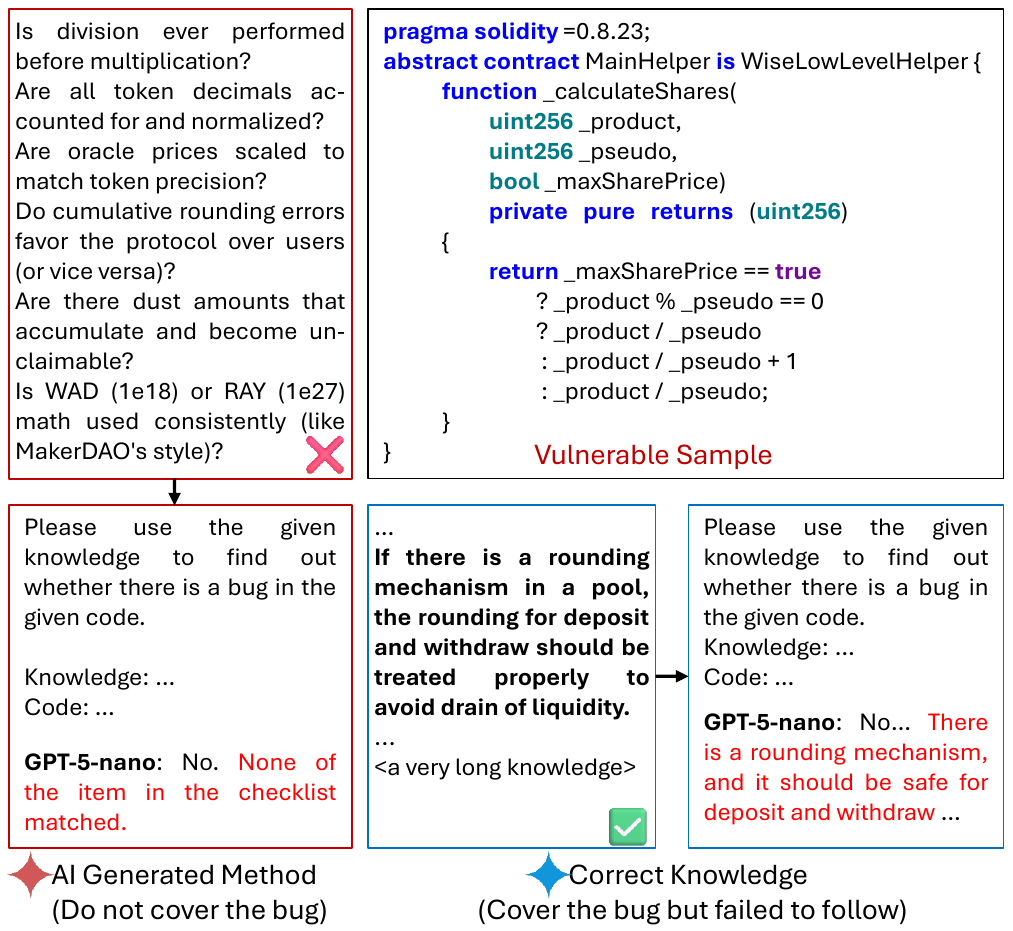}
    \caption{A precision issue vulnerability analyzed with two detection methods. Red Box in the left: An AI-generated method with insufficient coverage leads GPT-5-nano to a FN. Blue Box in the bottom right: A correct method is provided, but GPT-5-nano fails to follow it, also producing a FN.}
    \label{fig:example}
\end{figure}

To concretely illustrate why simply prompting an LLM fails to overcome these bottlenecks, consider a real-world smart contract snippet with an exploitable precision loss vulnerability (\Cref{fig:example}, top right in the black box).
Specifically, \texttt{\_calculateShares} improperly applies a ceiling operation during deposits: if the division yields a remainder, it grants an extra share (\texttt{\_product / \_pseudo + 1}). An attacker can exploit this asymmetric rounding by inflating the pool's assets (the denominator) via a direct donation, and then executing calculated micro-deposits to repeatedly trigger this remainder condition. This artificially inflates their share balance at negligible cost, ultimately enabling them to drain the protocol's liquidity. However, when an LLM attempts to detect this vulnerability, the process can fail in two distinct ways, mapping directly to our identified risks:

\begin{description}
    \item[Risk 1:] \textbf{The generated knowledge may be flawed or insufficient.}
    As shown in \Cref{fig:example} (red box, left), an AI-generated method can easily miss critical domain-specific patterns. While it checks generic rules (e.g., division before multiplication), it completely overlooks the specific rounding mechanism required for safe pool deposits. Consequently, when GPT-5-nano~\cite{gpt5nano2026} applies this method, it reports the contract as safe. This results in a false negative (FN) caused entirely by the knowledge gap, despite the model's correct reasoning over the provided rules.

    \item[Risk 2:] \textbf{The LLM may fail to strictly adhere to the detection knowledge.}
    Even when provided with a correct, comprehensive method (\Cref{fig:example}, blue box, bottom right), LLMs frequently struggle to execute multi-step logic faithfully~\cite{sun2024gptscan}. Given the correct method, GPT-5-nano skips required intermediate checks and reverts to its internal priors, yielding another FN. 
    Since an LLM's internal beliefs often override its own explicitly prompted reasoning, such as chain-of-thought~\cite{wei2022chain} or symbolic chain-of-thought~\cite{xu2024faithful}, relying on the model's autonomous execution is inherently unreliable~\cite{boppana2026reasoningtheaterdisentanglingmodel}.
    This necessitates a more deterministic execution mechanism beyond the model's own reasoning.
\end{description}

Together, these two risks reveal a fundamental challenge: if the generated knowledge is flawed or insufficient, even faithful execution produces wrong results; and if the LLM does not faithfully follow the knowledge, even correct knowledge leads to wrong results.
Neither problem can mask the other, and both must be solved simultaneously for reliable detection.

To overcome these risks and build a robust, controllable, and accurate vulnerability detection method, there are three key challenges to be addressed.
The severe scarcity of labeled data inherently dictates the first challenge (C1). Furthermore, as demonstrated by Risk 2, reliable detection demands that the LLM faithfully execute the provided logic rather than reverting to its unpredictable prior biases, driving the second challenge (C2). Finally, to circumvent the manual rule engineering bottleneck and the pitfalls of statically generated, flawed methods (Risk 1), the system must be capable of dynamically generating and evolving its detection knowledge, leading to the third challenge (C3).

\begin{figure*}
    \centering
    \includegraphics[width=0.87\linewidth]{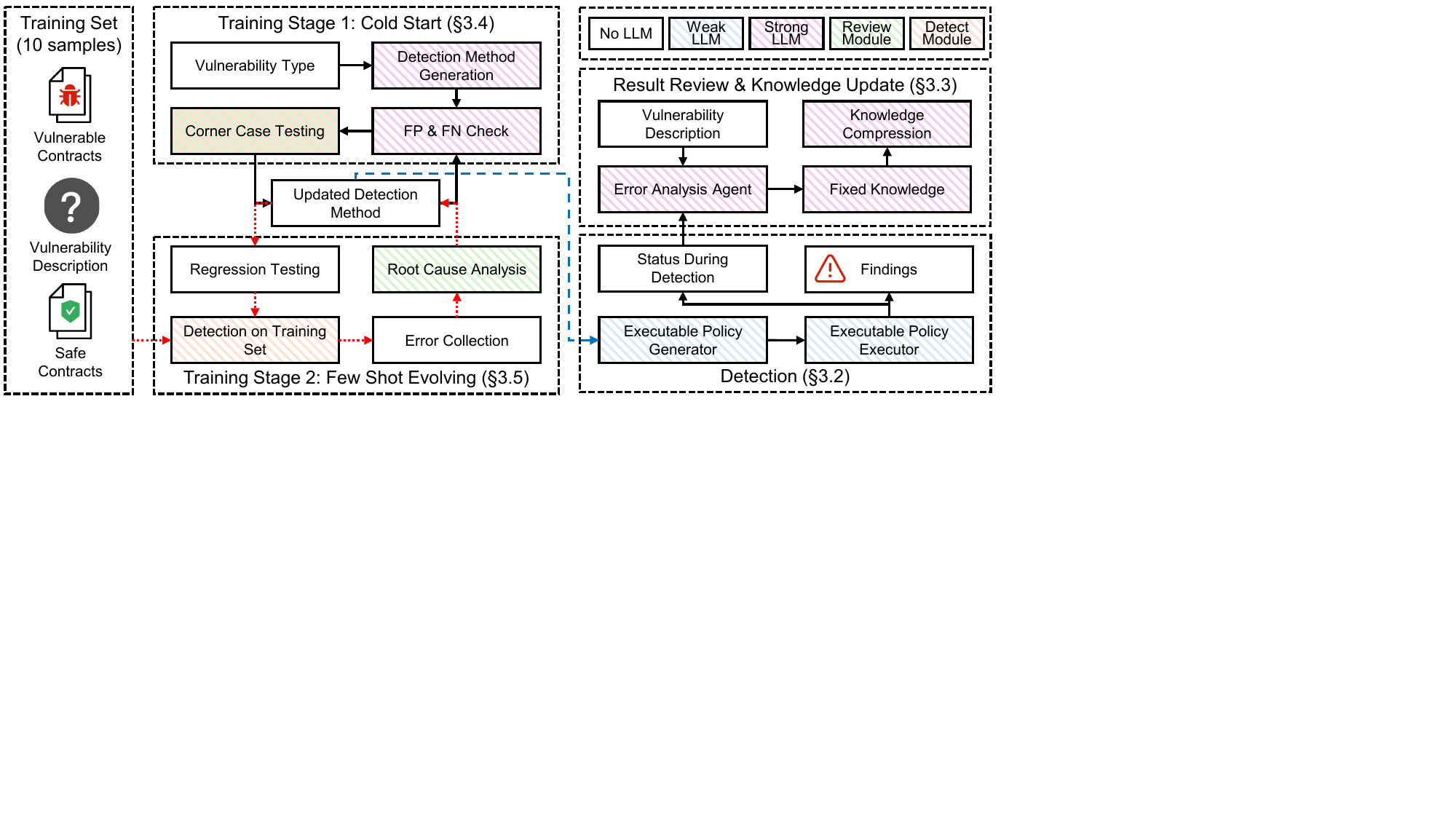}
    \caption{An overview of \tool.}
    \label{fig:overview}
\end{figure*}

\section{\tool}
\label{sec:method}

\tool frames vulnerability detection not merely as an inference task, but as an automated lifecycle management process for \textit{procedural detection knowledge}.
Rather than updating model parameters or relying on disposable reasoning traces, \tool represents detection expertise as a structured natural-language rule (a step-by-step procedure that instructs a detector how to identify a vulnerability in a smart contract) and automatically generates and refines this rule with only a few labeled samples, entirely eliminating the need for manual rule engineering.
To achieve this, \tool employs two LLMs with distinct roles: a \textit{supervisor} that analyzes detection errors and updates the rule, and a \textit{detector} that applies the current rule to target contracts to produce verdicts.
This decoupling allows the supervisor to be a capable but expensive model used only during training, while the detector can be a lightweight model deployed at inference time.

\subsection{Pipeline Overview}
\label{sec:overview}

The architecture of \tool is conceptually inspired by software engineering practices such as continuous integration (CI) and regression testing, mapped onto the semantic space of LLM reasoning. 
While it shares conceptual similarities with the reinforcement learning, \tool operates entirely in the semantic space: the \textit{detection rule} in natural language acts as the policy, detection errors on labeled contracts serve as the reward signals, and rule revisions serve as the policy updates. 
Throughout this process, no model parameters are modified.
This design yields three components, each targeting one of the challenges identified in~\cref{sec:motivation} (as shown in~\Cref{fig:overview}):

\noindent\textbf{\textit{Runtime}} (\cref{sec:runtime}) addresses \textbf{C2} (Logical Adherence) by exploiting the core capability dichotomy of modern LLMs: while they exhibit exceptional localized semantic comprehension (e.g., parsing the intent of a specific smart contract function), they notoriously struggle with long-horizon, rigid logical adherence~\cite{sun2024gptscan}.
It introduces a strict \textit{Inversion of Control} (IoC) mechanism. Instead of treating code generation as a disposable reasoning trace (as seen in Program-of-Thought), \tool compiles each detection rule into an \textit{Executable Policy} (EP), a deterministic programmatic detection plan executed in a sandboxed environment. The EP serves as a deterministic controller that strictly enforces the structural control flow of the detection process. Whenever localized semantic reasoning is required, the EP issues queries to the LLM via tightly-scoped \textit{Semantic Primitives}. This architectural design decouples the rigid execution path from the probabilistic reasoning, relying on the runtime for structural adherence while utilizing the LLM solely as an on-demand semantic oracle.
The execution environment simultaneously logs every reasoning step as a trace, transforming a binary wrong-prediction signal into a precise diagnosis.
Operating alongside the \textit{Runtime} is the \textbf{Result Review \& Knowledge Update} mechanism (\cref{sec:review}): triggered by any detection error, this independent component consumes the execution trace to identify the exact logical failure and revise the rule accordingly, essentially functioning as an automated semantic debugger.

\noindent \textbf{\textit{Cold Start}} (\cref{sec:cold-start}) addresses \textbf{C3} (Automated Extensibility) by bootstrapping a validated initial detection rule for a new vulnerability type without labeled contracts.
\tool instead elicits an initial rule from the LLM's pretraining security knowledge and stress-tests it against LLM-synthesized corner cases, all without any labeled data. This phase effectively acts as a semantic fuzzing process, producing a validated starting rule before a single real-world contract is seen.

\noindent\textbf{\textit{Few-Shot Evolving}} (\cref{sec:few-shot}) addresses \textbf{C1} (Data Efficiency) by iteratively refining the detection rule based on real-world contract feedback.
\tool applies the rule to real-world contracts and revises it based on execution feedback, requiring only a few labeled samples per vulnerability type and no parameter updates.
Crucially, this phase incorporates strict regression bounds, mirroring CI pipelines in software development.

\subsection{\textit{Runtime}}
\label{sec:runtime}

A natural-language detection rule captures vulnerability patterns in general, contract-agnostic terms. Applying it to a specific contract requires instantiating those semantic conditions against concrete functions and variables.
The \textit{Runtime} handles this by compiling the rule into an EP that traverses the target contract's structure and poses the rule's semantic questions in the context of specific code elements.
This keeps the detection knowledge portable across contracts.
Furthermore, enforcing a strict IoC, where verdicts are derived deterministically from the EP's execution logic rather than from unconstrained LLM generation, prevents the detector from reverting to its prior biases, directly addressing Risk~2 in~\cref{sec:motivation}.

\begin{figure}[t]
    \centering
    \includegraphics[width=.48\textwidth]{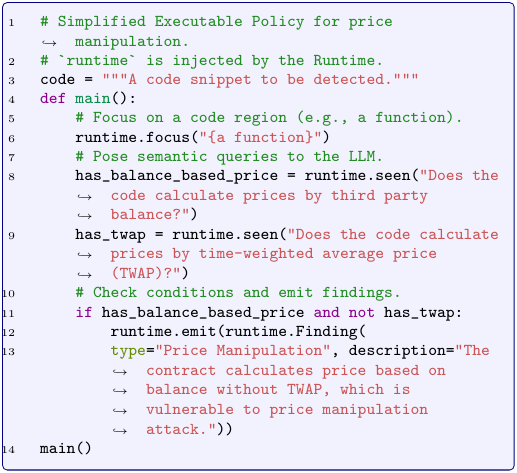}
    \caption{An example of an Executable Policy.}
    \label{lst:plan-example}
\end{figure}

\noindent\textbf{EP Generation \& Execution.}
Given a detection rule and a target contract, the LLM generates an EP (\Cref{lst:plan-example}) that iterates over the contract's functions, instantiates the rule's semantic checks against specific code elements, and emits findings when the conditions are met. 
Crucially, our EP differs from prior code-as-reasoning paradigms: unlike Program-of-Thought (PoT)~\cite{chen2023program}, which has the LLM author a disposable program that fully determines the computation, or symbolic chain-of-thought~\cite{xu2024faithful}, which generates a symbolic representation and then lets the LLM simulate the reasoning over it, our EP employs a \textit{hybrid execution model}. 
It interleaves deterministic programmatic control flow with on-demand LLM semantic queries. 
Rather than replacing the LLM, the EP acts as a structural manager that dictates exactly \textit{when} and \textit{what} to ask the LLM, keeping the semantic reasoning strictly bound to the prescribed logic.

\noindent\textbf{\textit{Runtime} Primitives.}
To operationalize the semantic-to-concrete translation within the generated EP, the \textit{Runtime} exposes three pre-defined primitives that serve as the strict execution interface between the deterministic controller and the LLM semantic oracle:

\noindent\underline{\textit{Focus}} (line~6).
This primitive narrows the analysis context to a specific contract or function by passing its source text to the \textit{Runtime}.
Subsequent semantic queries are evaluated against this focused snippet, with the full contract available as background context.

\noindent\underline{\textit{Seen}} (lines~8--9).
This primitive poses a natural-language question about the currently focused code to an LLM, receiving a boolean answer.
Crucially, all semantic evaluations must route exclusively through this primitive. The EP is prohibited from using string matching, regular expressions, or any syntactic pattern lookup.
This constraint keeps the detection logic at the semantic level described by the rule, and makes every reasoning step attributable to a specific natural-language question rather than to syntactic heuristics.

\noindent\underline{\textit{Emit}} (lines~12--13).
When the sequence of boolean responses satisfies the rule's vulnerability conditions, this function triggers a finding. 
It explicitly records the vulnerability type, the precise location, and a descriptive justification.

\noindent\textbf{Execution Constraints \& Logging.}
To ensure reliability, the EP executes under strict runtime constraints. First, each semantic query is issued as a stateless, independent request, ensuring the outcome is determined solely by the rule's logic rather than accumulated conversational context or prior attention-window hallucinations. Second, the \textit{Runtime} maintains an exhaustive log of the code snippets, semantic questions, boolean answers, logical branches, and emitted findings. This transforms opaque failures into structured traces for precise diagnosis, serving as the primary artifact for the \textbf{Result Review \& Knowledge Update} mechanism (\cref{sec:review}) when an error occurs.

\subsection{Result Review \& Knowledge Update}
\label{sec:review}

The \textbf{Result Review \& Knowledge Update} mechanism is shared by both \textit{Cold Start} and \textit{Few-Shot Evolving}.
In contrast to naive trial-and-error approaches where a wrong prediction yields a sparse, binary failure signal, \tool leverages the hybrid execution model to perform semantic debugging.
The structured execution trace produced by the \textit{Runtime} provides dense, step-level feedback, allowing the supervisor to attribute the failure to a specific reasoning step and revise only the relevant part of the rule.
The mechanism operates through a three-step automated lifecycle.

\noindent\textbf{Step 1: Root cause analysis.}
Upon encountering an error, the mechanism analyzes the full diagnostic context: the current detection rule, the generated EP, the contract source code, and the step-by-step trace of semantic queries and boolean answers.
By contrasting the ground truth against the detector's actual execution path, it identifies the precise logical gap, either an omission of a necessary check or a misclassification of a benign pattern as vulnerable.
To ensure high-precision localization, contextual anchors are dynamically injected depending on the error source.
When the error originates from the Corner Case-Based Evolving stage, the known gap description produced during corner case generation is provided to prevent spurious explanations (\cref{sec:cold-start}).
When the error occurs on a real vulnerable contract during \textit{Few-Shot Evolving}, the corresponding attack event report is injected, supplying the supervisor with a ground-truth description of the practical exploitation mechanics (\cref{sec:few-shot}).

\noindent\textbf{Step 2: Knowledge update.}
Based on the root cause analysis, the detection rule is revised to address the identified flaw.
To maintain the structural integrity of the procedural knowledge, \tool enforces a full-rewrite discipline rather than incremental, localized edits.
The supervisor must generate a complete, coherent rule that incorporates the specific patch while perfectly preserving the unaffected logic from the previous version.
This strict requirement prevents the classic failure modes of automated patching: inserting isolated clauses often breaks the logical flow of the rule, and blind semantic merging can silently corrupt previously verified checks without triggering immediate awareness.

\noindent\textbf{Step 3: Knowledge compression.}
Each update round risks introducing redundancy, as new clauses addressing edge cases accumulate alongside older, overlapping rules.
To prevent this accumulation of technical debt, \tool executes a mandatory refactoring step after every update. This process distills the rule into a compact, non-redundant state, merging overlapping conditions and eliminating verbosity while strictly preserving the logical boundaries of the detection behavior.
This guarantees that the resulting EPs remain structurally concise, preventing execution complexity from scaling with update rounds and maintaining inference stability.

\noindent\textbf{Batch update.}
When multiple errors occur within a single evaluation epoch, applying fixes sequentially often leads to oscillating updates where fixing one edge case introduces a regression in another.
To ensure monotonic improvement, \tool aggregates the root cause analyses for all errors and computes a global, batch-resolved update.
This CI-inspired conflict resolution ensures that every iteration produces a globally coherent, regression-free advancement of the detection knowledge, stabilizing the evolution process.

\subsection{\textit{Cold Start}}
\label{sec:cold-start}

\textit{Cold Start} produces a validated, robust baseline detection rule before the system observes a single labeled contract.
If the rule were initialized directly from the LLM's intrinsic latent knowledge and passed straight to \textit{Few-Shot Evolving}, the small number of labeled examples would bear the entire burden of shaping the rule. This risks severe overfitting, where the rule memorizes the idiosyncrasies of specific training contracts rather than capturing the underlying vulnerability invariants.
\textit{Cold Start} mitigates this vulnerability by first eliciting a broad initial heuristic (\textit{Knowledge Initialization}), and subsequently subjecting it to an iterative self-testing loop using auto-generated corner cases (\textit{Corner Case-Based Evolving}). This ensures the policy enters the empirical grounding phase with broad coverage and structural generality, providing a validated initial state that substantially shortens the subsequent optimization compared to starting from scratch.

\subsubsection{Knowledge Initialization}

Knowledge Initialization elicits an initial detection rule through a two-round dialogue.
The LLM is first instructed to reason freely about the vulnerability, constrained to express detection knowledge exclusively as statically observable source-code patterns (structural properties, statement ordering, and the presence or absence of modifiers and function calls), since \tool operates as a source-code-level detector and cannot reference runtime behaviors, transaction sequences, or off-chain conditions.
Free-form reasoning precedes structured extraction, following the insight from chain-of-thought prompting~\cite{nong2024chain,xu2024preemptive,huang2026towards,mao2025towards} that forcing immediate structured output suppresses the intermediate reasoning steps that improve output quality.
A follow-up turn then extracts the detection rule as a self-contained description, which is persisted as the starting point for corner case refinement.

\subsubsection{Corner Case-Based Evolving}

With an initial rule in hand, \tool refines it through an iterative self-testing loop driven by two LLM roles: a \textit{reviewer} that reads the current rule and generates synthetic Solidity corner cases to challenge its logic, and a \textit{detector} that applies the rule to those contracts and reports findings.
The loop exploits the reviewer's ability to reason about the rule's weaknesses and construct targeted contracts that expose them, entirely within the LLM's own reasoning space.
The rule must pass \textit{both} an FN probe and an False Positive (FP) probe before advancing to \textit{Few-Shot Evolving}; failure on either triggers another round of corner case generation and rule revision via the Result Review mechanism.

\noindent\textbf{FN probe.}
To expose potential FNs in unseen contracts, the reviewer executes four steps:
(1) identify which vulnerability patterns the rule would miss and why, without generating any code;
(2) synthesize a single Solidity contract containing a genuine vulnerability that exploits the identified gap;
(3) produce a plain-English explanation of why the contract is vulnerable, serving as an oracle independent of the detection rule's perspective;
(4) run the detector on the synthesized contract and, if the expected FN materializes, record the gap description and invoke Result Review to revise the rule to cover the newly exposed case.

\noindent\textbf{FP probe.}
The FP probe is structurally symmetric: the reviewer judges whether the rule would produce FPs on safe contracts, identifies which patterns would be incorrectly flagged or over-fitted, synthesizes one such safe contract, explains why it is benign, runs detection and triggers a rule revision if errors exist.

\noindent\textbf{Context compression.}
Each iteration accumulates generated Solidity code and multi-step reasoning in the agent's dialogue history.
To prevent context-window exhaustion and avoid re-debating already-resolved gaps, \tool replaces the full message history at the end of every iteration with a concise lessons-learned summary that preserves awareness of which gaps have been identified and resolved while discarding verbatim code that is no longer needed.

The \textit{Cold Start} module outputs a corner case-validated rule that has passed both the FN and FP probes on LLM-synthesized contracts.
This rule is subsequently handed to the \textit{Few-Shot Evolving} module, which anchors it to real-world labeled contracts.

\subsection{\textit{Few-Shot Evolving}}
\label{sec:few-shot}

Despite surviving synthetic corner-case validation, the \textit{Cold Start} rule is validated purely against synthetic distributions. It may still lack awareness of complex, multi-contract interactions or emergent exploit patterns unique to real-world deployments.
While traditional paradigms attempt to bridge this gap via parameter fine-tuning, such approaches are highly data-hungry and impractical given the severe scarcity of labeled on-chain exploits.
\textit{Few-Shot Evolving} resolves this by acting as an empirical grounding pipeline. It aligns the rule with real-world architectural patterns through targeted semantic revisions, requiring only a minimal set of labeled examples without modifying any underlying neural weights.

\noindent\textbf{Iterative CI/CD Lifecycle.}
Given the labeled training set, \tool refines the rule over multiple epochs, each consisting of three phases.

\noindent\textit{Phase 1: Detection \& Evaluation.}
\tool uses the \textit{Runtime} with an LLM for detection on all labeled contracts using the current rule, classifying each outcome as a true positive, true negative, false positive, or false negative.
If no errors are found, \tool terminates early: the rule has converged on the current training set.

\noindent\textit{Phase 2: Error Analysis and Unified Knowledge Update.}
Each detected error is analyzed in an independent session by the Result Review mechanism (\cref{sec:review}), using the current rule, the generated detection plan, the contract source, and the full execution trace.
Once all analyses are complete, \tool consolidates the root cause analyses and applies a single batch update to the rule, preventing the common failure mode where fixing one error in isolation introduces a regression on another. Compression is applied after the update.

\noindent\textit{Phase 3: Regression Check.}
Before accepting the updated rule, \tool re-runs the \textit{Runtime} on all contracts that produced errors in Phase 1.
The update is accepted only if the number of remaining errors is strictly fewer than before; otherwise the candidate rule is discarded and the previous version is retained.
This monotonicity guarantee ensures that each epoch either improves the rule or leaves it unchanged, preventing knowledge drift over successive updates.

After this stage, the detection rule has been validated against real-world contracts rather than purely LLM-synthesized ones, mitigating the risk that the rule reflects the model's internal assumptions rather than patterns actually observed in practice.

\noindent\textbf{Why few labeled examples suffice.}
A natural critique of any few-shot approach is whether such a minimal dataset can yield statistically meaningful generalizations. In traditional deep learning or parameter fine-tuning, learning from a handful of samples is intractable because the model must inductively extract underlying features from sparse, binary labels. \tool fundamentally bypasses this limitation by shifting the paradigm from \textit{inductive learning} to \textit{abductive semantic debugging}, leveraging two key principles:

\noindent\textit{(1) Pre-trained Semantic Abstraction.} LLMs enter this stage already possessing vast, pre-trained latent representations of security audits, vulnerabilities, and code semantics. Thus, the few-shot examples are not used to teach the LLM \textit{what} a vulnerability is from scratch. Instead, they act as specific examples that help the model match its general knowledge to real-world cases~\cite{zhou2024large,shahriar20255gpt,pearce2023examining,yang2025hybrid}.

\noindent\textit{(2) High Information Density via Execution Traces.} More importantly, \tool extracts orders of magnitude more information from a single failed sample than traditional methods. When a prediction fails, \tool does not merely return a binary ``incorrect'' signal; it yields a dense, step-by-step execution telemetry trace. Just as a human software engineer does not require thousands of failed test cases to fix a logical bug, a single failing test case with a clear stack trace is sufficient for fault localization, the supervisor LLM utilizes this dense trace to pinpoint the exact logical misstep. This mechanistic transparency allows a single labeled sample to drive a deterministic, highly targeted patch to the rule, rendering massive datasets unnecessary.

\section{Evaluation}

In this section, we evaluate \tool on five prevalent vulnerability types, covering both traditional vulnerabilities and logic bugs: Price Manipulation, Access Control, Insufficient Validation, Precision Issues, and Reentrancy.
We aim to answer the following research questions (RQs):
\begin{description}
    \item[RQ1: Effectiveness of \tool.] How effective is \tool in detecting real-world vulnerabilities across different types, and how does it compare to state-of-the-art baselines?
    \item[RQ2: Generalizability Across Models.] How is the performance of \tool affected by the model, and how does the performance change with different LLMs used in detectors?
    \item[RQ3: Ablation Study.] How does each component of \tool contribute to its overall performance?
    \item[RQ4: Cost Analysis.] How is the cost of \tool in terms of token cost during training and inference?
\end{description}

\subsection{Evaluation Setup}
The dataset, model selection, baselines, and training configuration used across all research questions are described below.

\subsubsection{Dataset}
In the evaluation, two datasets are used: (1) vulnerabilities collected from DeFiHackLab's collection of attack events~\cite{defihacks2026} %
, and (2) safe contracts collected from verified contracts on Etherscan.
Both datasets have been used by previous works~\cite{sun2024gptscan}. %
Since DeFiHackLab's collection of attack events is updated continuously, we manually collected a snapshot from the website.
We applied the following criteria to filter the collected data:
(1) the vulnerability type must be clearly labeled (e.g., price manipulation), excluding ambiguous categories such as ``logic flaw'';
(2) each vulnerability type must have at least 10 samples, with 5 reserved for training and 5 for testing;
(3) the source code of the vulnerable contract must be publicly available, as a number of attack events in the collection do not include verifiable on-chain source code;
(4) vulnerability type labels follow the original classification provided by DeFiHackLabs without further merging or splitting.
For the safe contract dataset, we are using the Top200 dataset from Sun~\etal~\cite{sun2024gptscan} %
and we randomly sample 50 of them for evaluation.
\Cref{tab:dataset} shows the details of the dataset.

\begin{table}[t]
    \centering
    \caption{The composition of vulnerability types versus benign samples within the dataset.}
    \label{tab:dataset}
    \resizebox{\linewidth}{!}{
    \begin{tabular}{lclc}
        \toprule
        \textbf{Vulnerability Type} & \textbf{\# Samples} & \textbf{Vulnerability Type} & \textbf{\# Samples} \\
        \midrule
        Price Manipulation  & 47                & Precision Issues    & 11 \\
        Access Control      & 33                & Reentrancy          & 27 \\
        Insufficient Validation & 41            & Non-Vulnerable      & 50 \\
        
        \midrule
        Total             & &  & 209 \\
        \bottomrule
    \end{tabular}}
\end{table}

\subsubsection{Model Selection}
In the evaluation, we are using GPT-5.2 as the supervising model for the training process, including both the \textit{Cold Start} and \textit{Few-Shot Evolving} stages.
For the student model used for detection, we are using a smaller and cheaper model, GPT-5-nano, to demonstrate the cost-effectiveness of \tool.
In RQ2, we will also evaluate the performance of \tool with different LLMs used in the detector, including Qwen3.5-9B~\cite{qwen3.5} and MiniMax-M2.5~\cite{minimaxm2.5}.
All of these models are provided by OpenRouter~\cite{openrouter}, %
with parameters set to default to avoid any bias in the evaluation.

\subsubsection{Baselines}
We compare \tool against four families of baselines. No single existing tool matches \tool's coverage, so each baseline is evaluated only on the vulnerability types it supports.
\textit{(i) Static analysis:} \underline{Slither}~\cite{feist2019slither}, which detects only reentrancy and access control.
\textit{(ii) LLM/DL-based detectors:} \underline{GPTScan}~\cite{sun2024gptscan}, a rule-based LLM method covering price manipulation, precision issues, insufficient validation, and access control (but not reentrancy); since its manually crafted rules only partially cover each type, we run it with both the paper's GPT-3.5-turbo and a GPT-5-nano variant (GPTScan (5-nano)). \underline{MANDO-GURU}~\cite{nguyen2022mando}, a graph-neural-network approach supporting access control, precision issues, and reentrancy. \underline{SAEL}~\cite{yu2025sael}, a fine-tuning-based Mixture-of-Experts method supporting only reentrancy, used via the authors' artifact with its full original training data (vs.\ five labeled samples for \tool). \underline{iAudit}~\cite{ma2025combining}, which combines fine-tuning with an LLM agent to produce detections with justifications.
\textit{(iii) Zero-shot LLMs:} \underline{GPT-5.2}~\cite{gpt522026} (\tool's supervising model) and \underline{GPT-5-nano}~\cite{gpt5nano2026} (its student model), without any training or evolving.
\textit{(iv) Coding agents:} \underline{Claude Code}~\cite{claudecode} (Claude Opus 4.8) and \underline{Codex}~\cite{codex} (GPT-5.5), each prompted to detect the five vulnerability types. %

\subsubsection{Training}
For training, we split the dataset into a training set and a testing set.
For training set, each vulnerability type contains only 5 vulnerable samples and 5 safe samples, which are used for the \textit{Cold Start} stage and the \textit{Few-Shot Evolving} stage.
In the \textit{Few-Shot Evolving} stage, we set the number of iterations to 5.

\begin{table*}[t]
    \centering
    \caption{Detection results across vulnerability types and methods (all values in \%).}
    \label{tab:main-results}
    \resizebox{\textwidth}{!}{
    \begin{tabular}{l | lrrr | lrrr | lrrr}
        \toprule
        \textbf{Vuln. Type}
            & \multicolumn{4}{c|}{\textbf{Our Method \& Ablations}}
            & \multicolumn{8}{c}{\textbf{Baselines}} \\
            & \textbf{Method} & \textbf{Prec.} & \textbf{Rec.} & \textbf{F1.}
            & \textbf{Method} & \textbf{Prec.} & \textbf{Rec.} & \textbf{F1.}
            & \textbf{Method} & \textbf{Prec.} & \textbf{Rec.} & \textbf{F1.} \\
        \midrule\midrule
        \multirow{5}{*}{Price Manip.}
            & \tool                  & \cv{82} & \cv{88} & \cv{85}
            & Slither                & \multicolumn{3}{c|}{\textit{Not Supported}}
            & MANDO-GURU             & \multicolumn{3}{c}{\textit{Not Supported}} \\
            & \tool (Qwen3.5-9B)     & \cv{79} & \cv{81} & \cv{80}
            & GPTScan (3.5)          & \cv{94} & \cv{36} & \cv{52}
            & iAudit                 & \cv{32} & \cv{50} & \cv{39} \\
            & \tool (MiniMax-M2.5)   & \cv{76} & \cv{83} & \cv{79}
            & GPTScan (5-nano)       & \cv{88} & \cv{36} & \cv{51}
            & SAEL                   & \multicolumn{3}{c}{\textit{Not Supported}} \\
            & \tool w/o Evolving     & \cv{53} & \cv{76} & \cv{63}
            & GPT-5.2 (0-shot)       & \cv{100} & \cv{62} & \cv{76}
            & Claude Code (Opus 4.8) & \cv{100} & \cv{64} & \cv{78} \\
            & \tool w/o Runtime      & \cv{100} & \cv{19} & \cv{32}
            & GPT-5-nano (0-shot)    & \cv{83} & \cv{57} & \cv{67}
            & Codex (GPT-5.5)        & \cv{100} & \cv{50} & \cv{67} \\
        \midrule
        \multirow{5}{*}{Access Control}
            & \tool                  & \cv{57} & \cv{93} & \cv{70}
            & Slither                & \cv{0} & \cv{0} & \cv{0}
            & MANDO-GURU\,(-64)$^1$  & \cv{22} & \cv{100} & \cv{36} \\
            & \tool (Qwen3.5-9B)     & \cv{55} & \cv{93} & \cv{69}
            & GPTScan (3.5)          & \cv{0} & \cv{0} & \cv{0}
            & iAudit                 & \cv{31} & \cv{71} & \cv{43} \\
            & \tool (MiniMax-M2.5)   & \cv{61} & \cv{87} & \cv{72}
            & GPTScan (5-nano)       & \cv{0} & \cv{0} & \cv{0}
            & SAEL                   & \multicolumn{3}{c}{\textit{Not Supported}} \\
            & \tool w/o Evolving     & \cv{43} & \cv{82} & \cv{57}
            & GPT-5.2 (0-shot)       & \cv{81} & \cv{60} & \cv{69}
            & Claude Code (Opus 4.8) & \cv{59} & \cv{79} & \cv{68} \\
            & \tool w/o Runtime      & \cv{73} & \cv{68} & \cv{70}
            & GPT-5-nano (0-shot)    & \cv{76} & \cv{57} & \cv{65}
            & Codex (GPT-5.5)        & \cv{84} & \cv{57} & \cv{68} \\
        \midrule
        \multirow{5}{*}{Insuff. Validation}
            & \tool                  & \cv{68} & \cv{78} & \cv{73}
            & Slither                & \multicolumn{3}{c|}{\textit{Not Supported}}
            & MANDO-GURU             & \multicolumn{3}{c}{\textit{Not Supported}} \\
            & \tool (Qwen3.5-9B)     & \cv{66} & \cv{70} & \cv{68}
            & GPTScan (3.5)          & \cv{100} & \cv{6} & \cv{11}
            & iAudit                 & \cv{42} & \cv{89} & \cv{57} \\
            & \tool (MiniMax-M2.5)   & \cv{63} & \cv{72} & \cv{67}
            & GPTScan (5-nano)       & \cv{100} & \cv{3} & \cv{5}
            & SAEL                   & \multicolumn{3}{c}{\textit{Not Supported}} \\
            & \tool w/o Evolving     & \cv{44} & \cv{83} & \cv{58}
            & GPT-5.2 (0-shot)       & \cv{72} & \cv{64} & \cv{68}
            & Claude Code (Opus 4.8) & \cv{60} & \cv{75} & \cv{67} \\
            & \tool w/o Runtime      & \cv{70} & \cv{67} & \cv{69}
            & GPT-5-nano (0-shot)    & \cv{53} & \cv{57} & \cv{50}
            & Codex (GPT-5.5)        & \cv{47} & \cv{72} & \cv{57} \\
        \midrule
        \multirow{5}{*}{Precision Issues}
            & \tool                  & \cv{40} & \cv{67} & \cv{50}
            & Slither                & \multicolumn{3}{c|}{\textit{Not Supported}}
            & MANDO-GURU\,(-44)      & \cv{0} & \cv{0} & \cv{0} \\
            & \tool (Qwen3.5-9B)     & \cv{50} & \cv{33} & \cv{40}
            & GPTScan (3.5)          & \cv{0} & \cv{0} & \cv{0}
            & iAudit                 & \cv{0} & \cv{0} & \cv{0} \\
            & \tool (MiniMax-M2.5)   & \cv{60} & \cv{50} & \cv{55}
            & GPTScan (5-nano)       & \cv{0} & \cv{0} & \cv{0}
            & SAEL                   & \multicolumn{3}{c}{\textit{Not Supported}} \\
            & \tool w/o Evolving     & \cv{9} & \cv{33} & \cv{14}
            & GPT-5.2 (0-shot)       & \cv{0} & \cv{0} & \cv{0}
            & Claude Code (Opus 4.8) & \cv{0} & \cv{0} & \cv{0} \\
            & \tool w/o Runtime      & \cv{0} & \cv{0} & \cv{0}
            & GPT-5-nano (0-shot)    & \cv{0} & \cv{0} & \cv{0}
            & Codex (GPT-5.5)        & \cv{20} & \cv{17} & \cv{18} \\
        \midrule
        \multirow{5}{*}{Reentrancy}
            & \tool                  & \cv{81} & \cv{77} & \cv{79}
            & Slither                & \cv{17} & \cv{5} & \cv{7}
            & MANDO-GURU\,(-44)      & \cv{22} & \cv{100} & \cv{36} \\
            & \tool (Qwen3.5-9B)     & \cv{79} & \cv{67} & \cv{74}
            & GPTScan (3.5)          & \multicolumn{3}{c|}{\textit{Not Supported}}
            & iAudit                 & \cv{17} & \cv{41} & \cv{24} \\
            & \tool (MiniMax-M2.5)   & \cv{66} & \cv{63} & \cv{64}
            & GPTScan (5-nano)       & \multicolumn{3}{c|}{\textit{Not Supported}}
            & SAEL                   & \cv{53} & \cv{73} & \cv{62} \\
            & \tool w/o Evolving     & \cv{36} & \cv{41} & \cv{38}
            & GPT-5.2 (0-shot)       & \cv{100} & \cv{32} & \cv{48}
            & Claude Code (Opus 4.8) & \cv{87} & \cv{59} & \cv{70} \\
            & \tool w/o Runtime      & \cv{86} & \cv{64} & \cv{74}
            & GPT-5-nano (0-shot)    & \cv{63} & \cv{55} & \cv{59}
            & Codex (GPT-5.5)        & \cv{100} & \cv{59} & \cv{74} \\
        \midrule\midrule
        \multirow{5}{*}{\textbf{Macro Avg.}}
            & \tool                  & \cv{66} & \cv{81} & \cv{71}
            & Slither                & \cv{8} & \cv{2} & \cv{4}
            & MANDO-GURU             & \cv{15} & \cv{67} & \cv{24} \\
            & \tool (Qwen3.5-9B)     & \cv{66} & \cv{69} & \cv{66}
            & GPTScan (3.5)          & \cv{48} & \cv{10} & \cv{16}
            & iAudit                 & \cv{24} & \cv{50} & \cv{33} \\
            & \tool (MiniMax-M2.5)   & \cv{65} & \cv{71} & \cv{67}
            & GPTScan (5-nano)       & \cv{47} & \cv{10} & \cv{14}
            & SAEL                   & \cv{53} & \cv{73} & \cv{62} \\
            & \tool w/o Evolving     & \cv{37} & \cv{63} & \cv{46}
            & GPT-5.2 (0-shot)       & \cv{71} & \cv{44} & \cv{52}
            & Claude Code (Opus 4.8) & \cv{61} & \cv{55} & \cv{57} \\
            & \tool w/o Runtime      & \cv{66} & \cv{44} & \cv{49}
            & GPT-5-nano (0-shot)    & \cv{55} & \cv{45} & \cv{48}
            & Codex (GPT-5.5)        & \cv{70} & \cv{51} & \cv{57} \\
        \bottomrule
    \end{tabular}}
    \vspace{0.5ex}
    \begin{minipage}{\textwidth}
    \footnotesize $^1$ \mbox{(-N)} after MANDO-GURU denotes the $N$ contracts on which it encountered a tool error (timeout, or unsupported newer Solidity compiler / multi-contract project) and produced no output; these contracts are excluded from its precision/recall computation (64 of 83 for access control, 44 of 77 for reentrancy, and 44 of 61 for precision issues).
    \end{minipage}
\end{table*}

\subsection{RQ1: Effectiveness of \tool}

In this RQ, we evaluate the effectiveness of \tool in detecting vulnerabilities in smart contracts, and compare it with the baselines and zero-shot performance of LLMs.
\Cref{tab:main-results} shows the detection results across all vulnerability types and methods, reporting the precision, recall and F1-score for each method.
One baseline warrants a note on scoring.
Because iAudit emits a vulnerability type rather than a plain binary verdict, we score it consistently with \tool's per-type evaluation: a detection counts as a true positive only if iAudit both flags the contract and assigns the correct type; a vulnerable contract that is missed or labeled with the wrong type is a false negative; and any safe contract flagged as vulnerable (of any type) is a false positive.
This type-aware criterion is, if anything, stricter than a purely binary vulnerable-vs-safe one, under which \tool's margin over iAudit would only widen.

\tool achieves a macro-average F1-score of 71\%, outperforming all baselines by a substantial margin: Slither (4\%), GPTScan (16\%), MANDO-GURU (24\%), iAudit (33\%), GPT-5-nano zero-shot (48\%), GPT-5.2 zero-shot (52\%), the coding agents Claude Code and Codex (57\% each), and SAEL (62\% on reentrancy only).
A notable characteristic of \tool is its high recall (81\%) at a reasonable precision (66\%), indicating it successfully identifies the majority of vulnerable contracts while keeping FPs at an acceptable level.

\noindent\textbf{Static analysis and deep-learning detectors.}
Slither and MANDO-GURU both reason over structural code properties and represent opposite failure modes.
Slither (macro F1-score=4\%) is confined to syntactic patterns: it cannot detect logic-level vulnerabilities such as price manipulation, insufficient validation, and precision issues, and even on its supported types it suffers from both FPs and FNs on real-world contracts.
MANDO-GURU (macro F1-score=24\% over its three supported types) instead over-detects, reaching 100\% recall but only 22\% precision on access control and reentrancy by flagging most contracts as vulnerable, and 0\% on precision issues.
It also produced tool errors on many contracts (64/83 access control, 44/77 reentrancy, 44/61 precision issues) due to unsupported newer Solidity compilers and multi-contract projects; we excluded these, which is the most favorable treatment for MANDO-GURU, since counting them as missed detections would only lower its recall further. Even so, it trails \tool by a wide margin.

\noindent\textbf{Rule-based and fine-tuning detectors.}
GPTScan, SAEL, and iAudit encode detection knowledge through manually crafted rules or supervised training, yet all fall well short of \tool.
GPTScan (macro F1-score=16\%) exhibits a striking precision-recall imbalance (48\% precision, 10\% recall): its hand-crafted rules cover only a subset of each type's patterns and miss most vulnerable contracts.
Notably, where its rules are relatively complete, such as price manipulation, it attains 94\% precision even with a weak GPT-3.5-turbo backend, confirming that well-crafted procedural knowledge with a structured checking procedure is highly effective, which is precisely what \tool constructs automatically.
The two fine-tuning-based methods remain weaker than \tool despite far larger training data: SAEL reaches F1-score=62\% on reentrancy, its only supported type, using its full original training set, whereas \tool attains 79\% using only five vulnerable examples.
iAudit (macro F1-score=33\%) over-detects, flagging nearly every contract as vulnerable, which yields low precision (17--42\%) and F1-scores of 39\%, 43\%, 57\%, and 24\% on price manipulation, access control, insufficient validation, and reentrancy, and 0\% on precision issues, trailing \tool on every evaluated type under the type-aware scoring described above.

\noindent\textbf{Zero-shot LLMs.}
The zero-shot baselines achieve macro-average F1-scores of 52\% (GPT-5.2) and 48\% (GPT-5-nano), performing reasonably on well-documented types such as price manipulation and access control, but failing on precision issues (F1-score=0\% for both).
\tool improves macro F1-score by 19\% over GPT-5.2 and 23\% over GPT-5-nano, demonstrating that the knowledge evolution process adds substantial value beyond the LLM's intrinsic capabilities.

\noindent\textbf{Coding agents.}
We further compare against two general-purpose coding agents, Claude Code (Opus 4.8) and Codex (GPT-5.5), which can autonomously inspect contract code over multiple steps.
Both reach a macro-average F1-score of 57\%, clearly above the zero-shot LLMs (48--52\%) owing to their agentic code-inspection ability, yet still well short of \tool's 71\%.
This gap is informative: even capable agentic systems, when applied directly, plateau because they lack reusable, vulnerability-specific detection knowledge.
\tool's advantage therefore stems from its knowledge-evolution mechanism rather than from the raw capability of LLM-based agents, and cannot be obtained simply by deploying a stronger general-purpose agent.

Looking at individual vulnerability types, \tool achieves the highest F1-score across all five types, including clear margins on price manipulation (85\%) and reentrancy (79\%).
On insufficient validation (F1-score=73\%), \tool improves over the best zero-shot baseline (GPT-5.2, F1-score=68\%) and far exceeds GPTScan (F1-score=11\%).
On precision issues (F1-score=50\%), \tool far outperforms every other method: only Codex achieves any non-zero result (18\%), while all remaining methods score F1-score=0\%, indicating that this vulnerability type requires specialized detection knowledge that cannot be derived from general pretraining alone.
On access control, \tool achieves F1-score=70\% with notably high recall (93\%), surpassing GPT-5.2 (F1-score=69\%) while both Slither and GPTScan score F1-score=0\% on this type.
Detecting access control flaws is challenging as it requires interpreting developer intent, but \tool achieves effective results.

Crucially, \tool's detector uses the lightweight GPT-5-nano rather than GPT-5.2. Despite this capacity gap, \tool achieves a 71\% macro F1-score, significantly outperforming GPT-5.2's 52\% in zero-shot settings. This 19-point lead shows that a structured execution framework can overcome raw model limitations, allowing a small, efficient model to outperform a much larger, unaided LLM.

\subsection{RQ2: Generalizability Across Models}

In this RQ, we evaluate whether the detection knowledge evolved by \tool can generalize to detector models other than the default GPT-5-nano.
After training with GPT-5.2 as the supervisor and GPT-5-nano as the detector, the resulting detection knowledge is transferred without modification to two alternative open-weight models: Qwen3.5-9B and MiniMax-M2.5.
The results are shown in~\Cref{tab:main-results}.
Note that due to intermittent instability of the OpenRouter API, including request timeouts and empty responses, 4 contracts were not evaluated for Qwen3.5-9B and 6 for MiniMax-M2.5; these samples are excluded from the reported metrics.

Both models achieve competitive results across all five vulnerability types.
\tool (Qwen3.5-9B) achieves a macro-average F1-score of 66\%, and \tool (MiniMax-M2.5) achieves 67\%, both substantially outperforming static analysis tools (Slither: 4\%), deep learning baselines (MANDO-GURU: 24\%), and LLM-based methods (GPTScan: 16\%), without any re-training or knowledge modification.
This demonstrates that the evolved detection knowledge is not overfit to a specific model's behavior and can be applied to different LLMs with reasonable effectiveness.

The performance gap between these models and \tool with GPT-5-nano (71\%) is likely because the detection knowledge is evolved using GPT-5-nano as the detector, so the knowledge refinement process is implicitly guided by GPT-5-nano's behavior and response patterns; some descriptions or phrasings in the knowledge may therefore be better suited to GPT-5-nano than to other models, leading to a modest performance gap when transferred.
Notably, MiniMax-M2.5 achieves F1-score=55\% on precision issues, which is higher than most baselines including GPT-5.2 zero-shot (F1-score=0\%), further confirming that the evolved knowledge provides substantial signal even when used with a different model.
Overall, the evolved knowledge generalizes across open-weight LLMs and lets them surpass the much larger GPT-5.2 in zero-shot mode (52\%) without any retraining.

\subsection{RQ3: Ablation Study}

In this RQ, we evaluate the contribution of each component of \tool to its overall performance.
\tool consists of two key components that can be independently ablated: the evolving process and the Runtime-based detector.
To evaluate the contribution of the evolving process, we remove both the Corner Case-Based Self-Testing stage and the Few-Shot Evolving stage, using only the initial detection method produced by Knowledge Initialization, denoted as \tool w/o Evolving.
To evaluate the contribution of the Runtime-based detector, we note that the Runtime's execution trace is a prerequisite for the training process, as it provides the structured feedback used by the Result Review and Knowledge Update module.
Therefore, \tool w/o Runtime is trained using the full system, and the Runtime is removed only at inference time, where it is replaced with a direct prompting approach that asks the LLM to classify each contract as vulnerable or not given the detection rule.
The results are shown in~\Cref{tab:main-results}.

Removing the evolving process (\tool w/o Evolving) reduces macro-average F1-score from 71\% to 46\%, a drop of 25 percentage points.
The degradation is primarily in precision, which falls from 66\% to 37\%, while recall remains relatively high at 63\%.
This reflects the inherent limitations of the LLM's built-in security knowledge: without iterative refinement through corner case testing and real-world labeled contracts, the initial rule captures many vulnerable patterns but also incorrectly flags a large number of safe contracts.
Continuous evolving is therefore essential for correcting the gaps and biases in the LLM's prior knowledge and aligning the detection rule with actual contract behavior.
The drop is most pronounced on reentrancy (F1-score: 79\% $\to$ 38\%) and precision issues (F1-score: 50\% $\to$ 14\%), vulnerability types whose real-world manifestations diverge most from the LLM's intrinsic assumptions.

Removing the Runtime-based detector (\tool w/o Runtime) reduces macro-average F1-score from 71\% to 49\%, a drop of 22 percentage points.
This ablation reveals three complementary findings.
First, without enforced execution, the LLM cannot reliably follow the detection rule, instead falling back on its own generation patterns.
Second, the behavior resembles a zero-shot detector: recall drops from 81\% to 44\% while precision remains at 66\%, confirming that the model effectively ignores the evolved rule and reverts to zero-shot-style logic.
Third, despite the absence of the \textit{Runtime}, \tool w/o \textit{Runtime} achieves higher precision than GPT-5-nano zero-shot (66\% vs.\ 55\%), demonstrating that the evolved knowledge retains value: it shifts the LLM's prior toward more accurate detection patterns compared to a zero-shot prompt.

Together, the two ablations confirm that both components make essential and complementary contributions: the evolving process refines the detection knowledge to match real-world patterns, while the Runtime ensures the knowledge is faithfully and systematically applied at inference time.

\subsection{RQ4: Cost Analysis}

In RQ4, we evaluate the economic cost of \tool. %
The time cost is not discussed here since it is highly dependent on throughput and latency of the underlying LLMs, which can vary significantly across different models and deployment settings, and may not be directly comparable.
For token cost, we calculate the average number of tokens used in the prompts during training and inference, divided by vulnerability type, and calculate the average token cost per thousand line of code (KLOC) for both training and inference.

\Cref{tab:cost} details the costs for the supervisor (Sup.) and detector (Det.). Overall, the economic cost of \tool is low and practical for real-world adoption.
Training for five vulnerability types consumes 2.67M/443K input/output tokens for the supervisor (GPT-5.2) and 541.85M/30.25M for the detector (GPT-5-nano). At official OpenAI rates (assuming no cache hits), the total training cost is \$50.06 (\$10.87 for the supervisor; \$39.19 for the detector). The inference phase, which utilizes only the detector to process 359 contracts (134 + 45 $\times$ 5), involves 306.62M input and 21.47M output tokens, resulting in a low maximum cost of \$23.70.

\begin{table}[t]
    \centering
    \caption{Token cost by vulnerability type and phase.}
    \label{tab:cost}
    \scalebox{0.93}{
    \begin{tabular}{llrrrr}
        \toprule
        \multirow{2}{*}{\textbf{Vulnerable Type}} & \multirow{2}{*}{\textbf{Phase}}
            & \multicolumn{2}{c}{\textbf{Input (K)}}
            & \multicolumn{2}{c}{\textbf{Output (K)}} \\
        \cmidrule(lr){3-4} \cmidrule(lr){5-6}
            & & \textbf{Sup.} & \textbf{Det.}
              & \textbf{Sup.} & \textbf{Det.} \\
        \midrule
        \multirow{2}{*}{Price Manipulation}
            & Train    & 567 & 64,551 & 100 & 5,570 \\
            & Infer & --$^{*}$ & 59,123 & -- & 4,444 \\
        \midrule
        \multirow{2}{*}{Access Control}
            & Train    & 715 & 98,897 & 116 & 5,950 \\
            & Infer & -- & 98,244 & -- & 5,459 \\
        \midrule
        \multirow{2}{*}{Insuff. Validation}
            & Train    & 604 & 44,367 & 99 & 5,437 \\
            & Infer & -- & 81,733 & -- & 5,985 \\
        \midrule
        \multirow{2}{*}{Precision Issues}
            & Train    & 442 & 227,416 & 57 & 6,234 \\
            & Infer & -- & 55,518 & -- & 2,738 \\
        \midrule
        \multirow{2}{*}{Reentrancy}
            & Train    & 339 & 106,623 & 71 & 7,064 \\
            & Infer & -- & 32,017 & -- & 2,839 \\
        \midrule
        \multirow{2}{*}{Total}
            & Train    & 2,667 & 541,854 & 443 & 30,255 \\
            & Infer & -- & 306,615 & -- & 21,465 \\
        \bottomrule
        \multicolumn{6}{l}{\footnotesize $^{*}$ Supervisor is not involved during inference; its token cost is marked as --.} \\
    \end{tabular}}
\end{table}

\section{Threats to Validity}

\noindent\textbf{Internal validity.}
The supervisor's root-cause analysis may occasionally misattribute trace errors, risking rule updates that fix one issue but introduce another. We mitigate this through strict epoch-level regression checks, which reject updates that cause net performance degradation.

\noindent\textbf{Construct validity.}
The \textit{Few-Shot Evolving} stage uses exactly five vulnerable and five safe samples per vulnerability type. To prevent selection bias, these are randomly drawn rather than hand-picked. A systematic analysis of how sample quantity and quality impact the evolved rules remains future work.

\noindent\textbf{External validity.}
First, regarding dataset scale: our evaluation totals 209 samples across five vulnerability types. While ``Precision Issues'' contains only 11 samples (leaving a statistically limited 6-sample test set), the other four categories feature substantially larger test sets that demonstrate consistent efficacy. This overall scale aligns with field norms for logic-bug evaluation (e.g., GPTScan~\cite{sun2024gptscan}, PropertyGPT~\cite{liu2025propertygpt}), which are fundamentally constrained by the intrinsic scarcity of verifiable exploits with public source code.

Second, regarding potential data leakage: historical exploits may exist in the LLMs' pre-training data. However, highly capable zero-shot baselines (e.g., GPT-5.2) fail to reliably detect these vulnerabilities on their own. This substantial performance gap confirms that \tool's gains stem from its synthesized procedural knowledge rather than mere memorization of leaked samples.

Finally, regarding generalizability: \tool evaluates Solidity contracts across five DeFi vulnerability types. While it cannot zero-shot detect entirely novel, undocumented vulnerabilities, its extensibility lies in bootstrapping effective detection for new types using minimal samples without manual engineering. Extending evaluation to cross-contract reasoning or other languages (like Vyper or Move) is currently infeasible due to data scarcity but remains a natural future direction. We have open-sourced our artifacts to facilitate such extensions.

\section{Related Works}

\noindent\textbf{Smart Contract Vulnerability Detection.}
Traditional static analysis~\cite{feist2019slither,Mythril,tsankov2018securify,torres2018osiris,mossberg2019manticore,luu2016making} and specialized vulnerability detectors~\cite{SliSE,Xue2020Reentrancy,ZibinZhengReentrancyStudy,ye2020clairvoyance,zhang2022reentrancy,reguard,song2025silence,Sailfish,Ghaleb_Rubin_Pattabiraman_2023,fang_modifier_issta_2023,Liu_2022_Finding,Zhong_2024_PrettySmart,lin2025actaint,liu2025have,kong2023defitainter,wu2023defiranger,mo2023toward} struggle with complex logic bugs due to their reliance on rigid predefined rules. 
To capture complex semantics, deep learning (DL) approaches~\cite{nguyen2022mando,li2024cobra,li2026novel} utilize neural representations but require massive labeled datasets that are inherently scarce in this domain. 
\tool bypasses both bottlenecks by evolving procedural knowledge from minimal samples, achieving generalized detection of logic bugs independent of rigid rules or data-hungry training.

\noindent\textbf{LLM for Code and Security Analysis.}
Recent works have applied LLMs to smart contract security via prompt engineering~\cite{chen2025chatgpt,david2023you}, advanced fine-tuning architectures (e.g., Smart-LLaMA-DPO~\cite{yu2025smart}, SAEL~\cite{yu2025sael}), or hybrid integrations with static tools (e.g., GPTScan~\cite{sun2024gptscan}, NumScout~\cite{chen2025numscout}, Ma~\etal~\cite{ma2025combining}). 
However, fine-tuned models remain static post-training, and hybrid methods inherit the rigidity of their underlying static analyzers. 
Unlike these approaches, \tool operates independently of legacy static tools and treats detection logic as an evolvable policy, continuously adapting to intricate logic vulnerabilities over time.

\noindent\textbf{Agentic Reasoning and Self-Refinement.}
Within the LLM agent community, self-refinement paradigms~\cite{madaan2023self,shinn2023reflexion} iteratively improve per-instance outputs based on LLM self-evaluation. 
Similarly, Program-of-Thoughts (PoT)~\cite{chen2023program} has the LLM author a disposable program that produces the answer, while symbolic chain-of-thought~\cite{xu2024faithful} generates a symbolic representation the LLM then simulates. In both, the LLM determines the entire computation in one shot, and the resulting reasoning is neither recorded as an auditable trace nor reused.
\tool departs from these one-shot paradigms in two dimensions.
First, rather than handing the whole task to the LLM, \tool executes each rule in a controlled, fully-logged environment that invokes the LLM only for localized semantic judgments (strict IoC), so every reasoning step is auditable.
Second, this trace is not discarded but serves as dense, ground-truth feedback for a persistent loop that refines reusable procedural knowledge, rather than relying on hallucination-prone self-assessment.

\section{Conclusion}
We presented \tool, an automated framework that reformulates smart contract vulnerability detection from a static inference task into a procedural knowledge evolution problem. 
Evaluations across five real-world vulnerability types show that \tool achieves a 71\% macro F1-score, substantially outperforming static analysis, deep learning-based, LLM-based methods, zero-shot LLM, and code agent baselines. 
Most notably, the evolved rules are portable: they enable lightweight, low-cost models to surpass much larger models. %
With ablations confirming the contribution of each component and a one-time evolution cost under \$50, \tool offers a practical, model-agnostic approach to smart contract vulnerability detection.

\bibliographystyle{IEEEtran}
\bibliography{ref}

\end{document}